\documentclass[conference,11pt]{IEEEtran}

\usepackage{array}
\usepackage[utf8]{inputenc}
\usepackage{amsmath,bm}
\usepackage{amsfonts}
\usepackage{amssymb}
\usepackage{amsthm}
\usepackage{graphicx,booktabs,longtable,xcolor}
\usepackage{algorithm,algorithmic}
\usepackage{booktabs}
\usepackage[printonlyused, nolist]{acronym}
\usepackage{trfsigns}
\usepackage{multirow} 
\usepackage{cite}
\usepackage{url}

\usepackage{tikz,pgfplots,pdflscape,pstricks}
\usetikzlibrary{arrows.meta,backgrounds,calc,decorations,patterns,positioning}
\usetikzlibrary{decorations.pathreplacing,angles,quotes,shapes,fit}
\usepgfplotslibrary{groupplots}
\newlength\fheight 
\newlength\fwidth 

\ifCLASSINFOpdf
\else
\fi

\definecolor{LMSred}{rgb}{0.80,0.20,0.20} 
\definecolor{LMSgray}{rgb}{0.60,0.60,0.60}
\definecolor{LMSCyan}{rgb}{0.60,1,1}
\definecolor{LMSMagenta}{rgb}{0.965,0.294,1}
\definecolor{LMSYellow}{rgb}{0.95,0.95,0.0}
\definecolor{LMSOrange}{rgb}{1,0.5,0}
\definecolor{LMSlightblue}{rgb}{0.5,0.8,1}
\definecolor{LMSlightred}{rgb}{1,0.294,0.294}
\definecolor{LMSblue}{rgb}{0.216,0.255,1}
\definecolor{LMSgreen}{rgb}{0.15,0.7,0.15}
\definecolor{LMSlightgreen}{rgb}{0.35,0.90,0.35}
\definecolor{LMSdarkgreen}{rgb}{0.01,0.5,0.01}

\begin{acronym}
\acro{ANN}{Aritifical Neural Network}
\acro{DFT}{Discrete Fourier Transform}
\acro{DCT}{Discrete Cosine Transform}
\acro{DNN}{Deep Neural Network}
\acro{MFCC}{Mel Frequency Cepstral Coefficient}
\acro{OR}{Outer Race}
\acro{IR}{Inner Race}
\acro{SVM}{Support Vector Machine}
\acro{MLP}{Multi-Layer perceptron}
\acro{DE}{Drive End}
\acro{NDE}{Non-Drive End}
\acro{TPR}{True Positive Rate}
\acro{TNR}{True Negative Rate}
\end{acronym}

\pgfplotsset{compat=1.18}
\begin{document}

\title{Airborne Sound Analysis for the Detection of Bearing Faults in Railway Vehicles with Real-World Data}

\author{Matthias Kreuzer${}^{1}$, David Schmidt${}^{2}$, Simon Wokusch${}^{2}$, Walter Kellermann${}^{1}$ \\
\textit{${}^{1}$Multimedia Communications and Signal Processing, FAU Erlangen-N\"urnberg, Germany} \\
\texttt{\{matthias.kreuzer, walter.kellermann\}@fau.de} \\
\textit{${}^{2}$ Siemens Mobility GmbH, N\"urnberg, Germany} \\
\texttt{\{schmidtdavid, simon.wokusch\}@siemens.com}} 

\maketitle

\begin{abstract}
In this paper, we address the challenging problem of detecting bearing faults in railway vehicles by analyzing acoustic signals recorded during regular operation. For this,  we introduce Mel Frequency Cepstral Coefficients (MFCCs) as features, which form the input to a simple  Multi-Layer Perceptron classifier. The proposed method is evaluated with real-world data that was obtained for state-of-the-art commuter railway vehicles in a measurement campaign. The experiments show that bearing faults can be reliably detected with the chosen MFCC features  even for bearing damages that were not included in training.
\end{abstract}
\vspace{0.2cm}


\section{Introduction}\label{sec:introduction}

Rolling element bearings are key components in a lot of rotating machinery, such as the induction motors of railway vehicles. Since bearings are exposed to wear, the maintenance of bearings is a non-negligible cost factor. Hence, reliable remote condition monitoring techniques are highly sought after.

To this end, the analysis of structure-borne sound using envelope analysis-, signal decomposition- and filtering techniques \cite{randall_rb_rolling_2011,peng2022} has been studied extensively and utilized for decades. More recently, end-to-end \ac{DNN}-based vibration fault detection approaches gained wide-spread popularity \cite{neupane2020,hamadache2019,zhang2020deep,ding2023} and became prevalent for the task of bearing fault detection. In parallel, the detection of faults based on the analysis of airborne sound has been attracting more and more attention in recent years \cite{yang2022,daraz2018,hou2022,zhang2018,yu2022}. The analysis of airborne sound analysis is highly promising since it comes with numerous advantages. Most importantly, airborne sound analysis  allows for an easy retroactive refitting of already existing (vehicle) platforms with condition monitoring equipment as microphones are less intrusive than acceleration sensors. Whereas acceleration sensors have to be mounted directly on the housing of the machine and as closely to the component that is to be monitored as possible, microphones can simply be placed in its proximity. Hence, the specifications regarding the placement of the sensors are more relaxed and, e.g., the motor or the gearbox of a railway vehicle does not have to be equipped with additional  mechanical interfaces as often necessary for acceleration sensors.  Moreover, a single sensor can be used to monitor multiple components. Finally, the sensing cost is usually significantly lower \cite{yun2020}.
Obviously airborne sound data could also be used to support structure-borne sound classification by providing additional information that can lead to more reliable classification results.

In \cite{Wu2022} a bearing fault feature extraction approach is proposed that combines Adaptive Variational Mode Decomposition (AVMD), an Improved Multiverse Optimization (IMVO) algorithm  and Maximum Correlated Kurtosis Deconvolution (MCKD) to subsequently identify fault features in the envelope spectrum of airborne sound.  Likewise, the Variational Mode Decomposition (VMD) is used for denoising in \cite{yan2022}. In \cite{yan2022}, which aims at the detection of cylinder misfires or blocked air inlets in a diesel engine, \acp{MFCC} are extracted from the denoised signals, which are then used to train a long short-term memory (LSTM) network which acts as the classifier.  
Another signal decomposition technique, i.e., the Fourier decomposition method (FMD), is applied in \cite{liu2016} for the task of classifying bearing faults. The time and envelope kurtosis are then extracted from the decomposed signals and subsequently used as features for training a random forest classification algorithm. Whereas the aforementioned approaches rely on extracting hand-crafted features, the task of identifying discriminant features was handed to a Stacked Auto-Encoder (SAE), which operates on the raw spectrograms of sound signals. 

However, all these approaches have been evaluated in laboratory environments using test benches. Additionally, the presented classifiers were not evaluated with unseen fault conditions, which they will most likely face in practical scenarios. Thus, in this paper we investigate the potential of airborne sound analysis for the detection of bearing faults in induction motors and gearboxes in a highly challenging scenario, i.e, a state-of-the-art commuter railway vehicle during normal operation. It is shown that \acp{MFCC} are particularly well-suited features and that they can be used in conjunction with a rather simple \acf{MLP} for the detection of bearing faults in a realistic condition monitoring scenario even when presented with previously unseen fault conditions.  

This article is structured as follows. In Sec.~\ref{sec:scenario} the considered scenario is described comprehensively: the railway vehicle is described in Sec.~\ref{sec:railway_vehicle}, the placement of the sensors is discussed in Sec.~\ref{sec:sensor_placement}, the bearing damages are described in Sec.~\ref{sec:bearing_damages}, and the data acquisition process is outlined in Sec.~\ref{sec:data_acquisition}. Thereupon the proposed bearing fault detection approach is presented in Sec.~\ref{sec:approach}, which is then evaluated in Sec.~\ref{sec:results} for seen and unseen damages in Sec.~\ref{sec:seen_damages} and Sec.~\ref{sec:unseen_damages}, respectively. Finally, conclusions are  drawn in Sec.~\ref{sec:conclusion}.

\section{Scenario}\label{sec:scenario}
In the following the investigated scenario is described, which includes a description of the railway vehicle, the sensor placement, the bearing damages, and the data acquisition process. 
\subsection{Railway Vehicle}\label{sec:railway_vehicle}
\begin{figure}[!ht]
    \centering
    \includegraphics[scale=1]{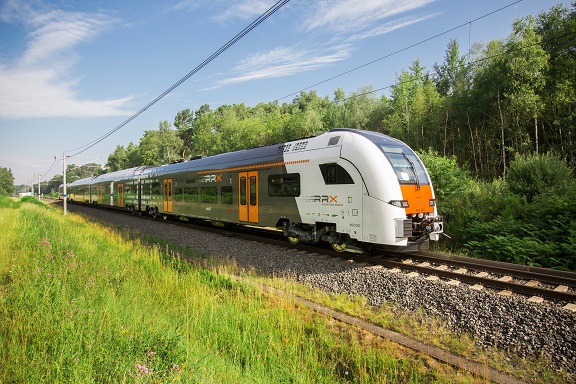}
    \caption{Photograph of the commuter rail vehicle of the type Desiro HC RRX that was equipped with measurement equippment to obtain the real-world data.}
    \label{fig:rrx}
\end{figure}
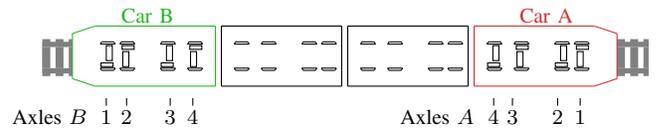
\begin{figure}[ht!]
    \centering
    \usetikzlibrary {shapes.geometric,calc} 

\def\engine#1#2{
\begin{scope}
\draw [draw=black] (#1,#2) rectangle ++ (0.4,1);
\end{scope}
}
\def\gearbox#1#2{
\begin{scope}
\draw [draw=black] (#1,#2) rectangle ++ (0.8,0.25);
\end{scope}
}
\def\wheeltop#1#2{
\begin{scope}[shift={(0,0)},rotate=0]
\draw [draw=black] (#1,#2) -- ++ (1,0);
\draw [draw=black] (#1,#2) -- ++ (0.1,0.15);
\draw [draw=black] ($(#1,#2) + (1,0)$) -- ++ (-0.1,0.15);
\draw [draw=black] ($(#1,#2) + (0.1,0.15)$) -- ($(#1,#2) + (0.9,0.15)$);
\end{scope}
}
\def\wheelbottom#1#2{
\begin{scope}[rotate=0]
\draw [draw=black] (#1,#2) -- ++ (1,0);
\draw [draw=black] (#1,#2) -- ++ (0.1,-0.15);
\draw [draw=black] ($(#1,#2) + (1,0)$) -- ++ (-0.1,-0.15);
\draw [draw=black] ($(#1,#2) + (0.1,-0.15)$) -- ($(#1,#2) + (0.9,-0.15)$);
\end{scope}
}
 \tikzstyle{track}=[
   postaction={draw=gray,densely dashed,line width=14pt},
   postaction={draw=gray, double distance=8pt,line width=2pt},
   postaction={draw=gray,densely dashed,line width=8pt},]

\begin{tikzpicture}[scale=0.2]
\node[text=LMSgreen] (waggona) at (-4.5,1) {\scriptsize Car B};
\node[text=LMSred] (waggonb) at (22.0,1) {\scriptsize Car A};
\draw[track] (-11.5,-1.8) -- (-9.5,-1.8);
\draw[track] (26.7,-1.8) -- (28.8,-1.8);

\node[] at (-11,-5.75) (axle) {\scriptsize Axles $B$};
\draw[] (-7.25,-4.5) -- node[pos=2.5] {\scriptsize  $1$}(-7.25,-5);
\draw[] (-5.9,-4.5) --  node[pos=2.5] {\scriptsize  $2$}(-5.9,-5);
\draw[] (-3,-4.5) --node[pos=2.5] {\scriptsize  $3$} (-3,-5);
\draw[] (-1.5,-4.5) -- node[pos=2.5] {\scriptsize  $4$}(-1.5,-5);

\begin{scope}[shift ={(25.75,0)}]
    \node[] at (-11,-5.75) (axle) {\scriptsize Axles $A$};
\draw[] (-7.25,-4.5) -- node[pos=2.5] {\scriptsize  $4$}(-7.25,-5);
\draw[] (-6,-4.5) --  node[pos=2.5] {\scriptsize  $3$}(-6,-5);
\draw[] (-3,-4.5) --node[pos=2.5] {\scriptsize  $2$} (-3,-5);
\draw[] (-1.5,-4.5) -- node[pos=2.5] {\scriptsize  $1$}(-1.5,-5);
\end{scope}
\begin{scope}[rotate=180,shift ={(0,-0.1)}]
\def\xend{8}
\def\yend{4}

\begin{scope}[shift ={(0,-0.2)}]
\draw[draw=LMSgreen] (0,0) -- (0,4);
\draw[draw=LMSgreen]  (0,0) -- (8,0);
\draw[draw=LMSgreen]  (0,4) -- (8,4);
\draw[draw=LMSgreen]  (9.5,3.5) -- (9.5,0.5);
\draw[draw=LMSgreen]  (8,4) -- (9.5,3.5);
\draw[draw=LMSgreen]  (8,0) -- (9.5,0.5);
\end{scope}
\begin{scope}
\engine{1.25}{1.5};
\gearbox{0.9}{1.1};
\wheeltop{0.85}{2.6};
\wheelbottom{0.85}{1};
\end{scope}
\begin{scope}[shift={(4.5,3.6)},rotate around={180:(0,0)}]
\engine{1.25}{1.5};
\gearbox{0.9}{1.1};
\wheeltop{0.85}{2.6};
\wheelbottom{0.85}{1};
\end{scope}
\begin{scope}[shift={(4.5,0)}]
\engine{1.25}{1.5};
\gearbox{0.9}{1.1};
\wheeltop{0.85}{2.6};
\wheelbottom{0.85}{1};
\end{scope}
\begin{scope}[shift={(8.5,3.6)},rotate around={180:(0,0)}]
\engine{1.25}{1.5};
\gearbox{0.9}{1.1};
\wheeltop{0.85}{2.6};
\wheelbottom{0.85}{1};
\end{scope}

\end{scope}
\begin{scope}[shift={(0.4,-3.5)}]
\def\xend{8}
\def\yend{4}

\begin{scope}[shift ={(0,-0.2)}]
\draw (0,0) -- (0,4);
\draw (0,0) -- (8,0);
\draw (0,4) -- (8,4);
\draw (8,4) -- (8,0);

\end{scope}
\begin{scope}

\wheeltop{0.85}{2.6};
\wheelbottom{0.85}{1};
\end{scope}
\begin{scope}[shift={(4.5,3.6)},rotate around={180:(0,0)}]

\wheeltop{0.85}{2.6};
\wheelbottom{0.85}{1};
\end{scope}
\begin{scope}[shift={(4.5,0)}]

\wheeltop{0.85}{2.6};
\wheelbottom{0.85}{1};
\end{scope}
\begin{scope}[shift={(8.5,3.6)},rotate around={180:(0,0)}]

\wheeltop{0.85}{2.6};
\wheelbottom{0.85}{1};
\end{scope}

\end{scope}

\begin{scope}[shift={(8.8,-3.5)}]
\def\xend{8}
\def\yend{4}

\begin{scope}[shift ={(0,-0.2)}]
\draw (0,0) -- (0,4);
\draw (0,0) -- (8,0);
\draw (0,4) -- (8,4);
\draw (8,4) -- (8,0);

\end{scope}
\begin{scope}

\wheeltop{0.85}{2.6};
\wheelbottom{0.85}{1};
\end{scope}
\begin{scope}[shift={(4.5,3.6)},rotate around={180:(0,0)}]

\wheeltop{0.85}{2.6};
\wheelbottom{0.85}{1};
\end{scope}
\begin{scope}[shift={(4.5,0)}]

\wheeltop{0.85}{2.6};
\wheelbottom{0.85}{1};
\end{scope}
\begin{scope}[shift={(8.5,3.6)},rotate around={180:(0,0)}]

\wheeltop{0.85}{2.6};
\wheelbottom{0.85}{1};
\end{scope}

\end{scope}

\begin{scope}[shift={(17.2,-3.5)}]
\def\xend{8}
\def\yend{4}

\begin{scope}[shift ={(0,-0.2)}]
\draw[draw=LMSred] (0,0) -- (0,4);
\draw[draw=LMSred]  (0,0) -- (8,0);
\draw[draw=LMSred]  (0,4) -- (8,4);
\draw[draw=LMSred]  (9.5,3.5) -- (9.5,0.5);
\draw[draw=LMSred] (8,4) -- (9.5,3.5);
\draw[draw=LMSred] (8,0) -- (9.5,0.5);
\end{scope}
\begin{scope}
\engine{1.25}{1.5};
\gearbox{0.9}{1.1};
\wheeltop{0.85}{2.6};
\wheelbottom{0.85}{1};
\end{scope}
\begin{scope}[shift={(4.5,3.6)},rotate around={180:(0,0)}]
\engine{1.25}{1.5};
\gearbox{0.9}{1.1};
\wheeltop{0.85}{2.6};
\wheelbottom{0.85}{1};
\end{scope}
\begin{scope}[shift={(4.5,0)}]
\engine{1.25}{1.5};
\gearbox{0.9}{1.1};
\wheeltop{0.85}{2.6};
\wheelbottom{0.85}{1};
\end{scope}
\begin{scope}[shift={(8.5,3.6)},rotate around={180:(0,0)}]
\engine{1.25}{1.5};
\gearbox{0.9}{1.1};
\wheeltop{0.85}{2.6};
\wheelbottom{0.85}{1};
\end{scope}

\end{scope}
\end{tikzpicture}
    \vspace{-0.75cm}
    \caption{Schematic view of the railway vehicle. Two cars, i.e., Car A and Car B, were equipped with condition monitoring equipment. The damaged bearings were
installed in Car A, whereas Car B was only equipped with healthy
bearings and hence serves as reference for the healthy state.}
    \label{fig:train_scheme}
\end{figure}

In order to obtain realistic bearing fault data, a state-of-the-art commuter railway vehicle of the type Desiro HC RRX (Rhein-Ruhr-Express) \cite{datasheet_rrx} as shown in Fig.~\ref{fig:rrx} was equipped with damaged bearings and a multitude of sensors was mounted on two cars. Two test trips with a duration of approximately 4 hours each were conducted. The railway vehicle was commuting between two cities to simulate normal operation.  Two of the four train cars (cf. Fig.~\ref{fig:train_scheme}), i.e., Car A and Car B, were equipped with measurement equipment, i.e, acceleration sensors, temperature sensors, microphones, etc. In this paper, only the microphone data is considered. A drivetrain, comprising a motor and a gearbox, is located at each of the four axles of a car (cf. Fig.~\ref{fig:train_scheme} and Fig.~\ref{fig:waggon_scheme}). The axles of Car B are denoted as $Bi$ with $i \in \{1,2,3,4\}$, while the axles of Car A are denoted as $Ai$ with $i \in \{1,2,3,4\}$.  At Car B, solely healthy bearings are installed and therefore the measurements of Car B serve as a reference for the healthy state of a bearing. However, Car A is equipped with damaged bearings, which will be described in Sec.~\ref{sec:bearing_damages}.

\subsection{Sensor Placement}\label{sec:sensor_placement}

\begin{figure}[ht!]
    \centering
    \input{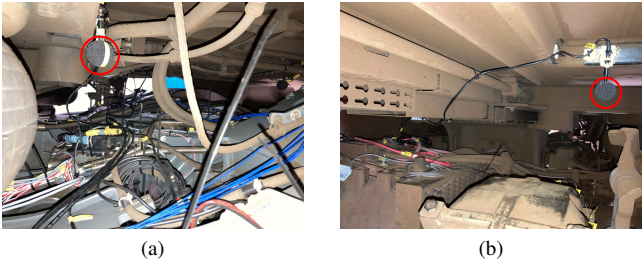}
    \vspace{-0.85cm}
    \caption{Placement of the microphones ($ \textcolor{red}{\circ}$) on the rail vehicle. A single microphone is placed above each drivetrain component by attaching it to the bottom of the railway car. Subfigure (a) depicts the microphone above the motor and Subfigure (b) shows the microphone above the gearbox.}
    \label{fig:mic_placement}
\end{figure}

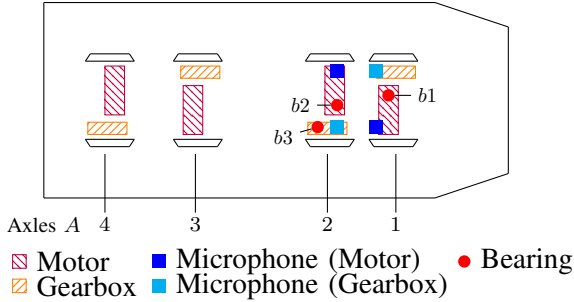
\begin{figure}[ht!]
 \begin{center}
     \usetikzlibrary {shapes.geometric,calc} 

\def\engine#1#2{
\begin{scope}
\draw [draw=purple,pattern={north west lines},pattern color=purple] (#1,#2) rectangle ++ (0.4,1);
\end{scope}
}
\def\gearbox#1#2{
\begin{scope}
\draw [draw=orange,pattern={north east lines},pattern color=orange] (#1,#2) rectangle ++ (0.8,0.25);
\end{scope}
}
\def\wheeltop#1#2{
\begin{scope}[shift={(0,0)},rotate=0]
\draw [draw=black] (#1,#2) -- ++ (1,0);
\draw [draw=black] (#1,#2) -- ++ (0.1,0.15);
\draw [draw=black] ($(#1,#2) + (1,0)$) -- ++ (-0.1,0.15);
\draw [draw=black] ($(#1,#2) + (0.1,0.15)$) -- ($(#1,#2) + (0.9,0.15)$);
\end{scope}
}
\def\wheelbottom#1#2{
\begin{scope}[rotate=0]
\draw [draw=black] (#1,#2) -- ++ (1,0);
\draw [draw=black] (#1,#2) -- ++ (0.1,-0.15);
\draw [draw=black] ($(#1,#2) + (1,0)$) -- ++ (-0.1,-0.15);
\draw [draw=black] ($(#1,#2) + (0.1,-0.15)$) -- ($(#1,#2) + (0.9,-0.15)$);
\end{scope}
}
 \tikzstyle{track}=[
   postaction={draw=black,densely dashed,line width=14pt},
   postaction={draw=black,double distance=8pt,line width=2pt},
   postaction={draw=black,densely dashed,line width=8pt},]
\begin{tikzpicture}[scale=0.65]
\def\xend{8}
\def\yend{4}

\begin{scope}[shift ={(0,-0.2)}]
\draw (0,0) -- (0,4);
\draw (0,0) -- (8,0);
\draw (0,4) -- (8,4);
\draw (9.5,3.5) -- (9.5,0.5);
\draw (8,4) -- (9.5,3.5);
\draw (8,0) -- (9.5,0.5);
\end{scope}
\begin{scope}
\engine{1.25}{1.5};
\gearbox{0.9}{1.1};
\wheeltop{0.85}{2.6};
\wheelbottom{0.85}{1};
\end{scope}
\begin{scope}[shift={(4.5,3.6)},rotate around={180:(0,0)}]
\engine{1.25}{1.5};
\gearbox{0.9}{1.1};
\wheeltop{0.85}{2.6};
\wheelbottom{0.85}{1};
\end{scope}
\begin{scope}[shift={(4.5,0)}]
\engine{1.25}{1.5};
\gearbox{0.9}{1.1};
\wheeltop{0.85}{2.6};
\wheelbottom{0.85}{1};
\end{scope}
\begin{scope}[shift={(8.5,3.6)},rotate around={180:(0,0)}]
\engine{1.25}{1.5};
\gearbox{0.9}{1.1};
\wheeltop{0.85}{2.6};
\wheelbottom{0.85}{1};
\end{scope}
\node[fill=red,circle,scale=0.45] (b2) at (7.05,1.9) {};
\node[scale=0.7,color=black] (b2_text) at (7.85,1.9) {$b1$};
\draw[] (b2) -- (b2_text);

\node[fill=red,circle,scale=0.45] (b1) at (6,1.7) {};
\node[scale=0.7,color=black] (b1_text) at (5.25,1.7) {$b2$};
\draw[scale=0.7,color=black] (b1) -- (b1_text);

\node[fill=red,circle,scale=0.45] (b3) at (5.6,1.25) {};
\node[scale=0.7,color=black] (b3_text) at (4.9,1.05) {$b3$};
\draw[] (b3) -- (b3_text);

\node[] at (0,-0.75) (axle) {\scriptsize Axles $A$};
\draw[] (1.25,0.7) -- node[pos=1.2] {\scriptsize  $4$}(1.25,-0.5);
\draw[] (3.1,0.7) --  node[pos=1.2] {\scriptsize  $3$}(3.1,-0.5);
\draw[] (5.8,0.7) --node[pos=1.2] {\scriptsize  $2$} (5.8,-0.5);
\draw[] (7.2,0.7) -- node[pos=1.2] {\scriptsize  $1$}(7.2,-0.5);
\node[fill=cyan,text=red,scale=0.7] (m1) at (6.8,2.4) {};
\node[fill=blue,text=red,scale=0.7] (m2) at (6.8,1.25) {};
\node[fill=blue,text=red,scale=0.7] (m3) at (6,2.4) {};
\node[fill=cyan,text=red,scale=0.7] (m4) at (6,1.25) {};

\node[fill=blue,text=black,scale=0.7,label=right:{Microphone (Motor)}] (mic) at (2.35,-1.5) {};
\node[fill=cyan,text=black,scale=0.7,label=right:{Microphone (Gearbox)}] (mic) at (2.35,-2) {};

\node[fill=red,text=black,label=right:{Bearing},circle,scale=0.45] (bearing) at (8.6,-1.5) {};
\node[pattern={north west lines},pattern color=purple,text=black,draw=purple,scale=0.7,label=right:{Motor}] (mic) at (-0.5,-1.5) {};
\node[pattern={north east lines},pattern color=orange,text=black,draw=orange,scale=0.7,label=right:{Gearbox}] (mic) at (-0.5,-2) {};
\end{tikzpicture}
 \end{center}
 \vspace{-0.5cm}
 \caption{Schematic view of Car A. The first two drivetrains, i.e, motor and gearbox,  are equipped with acoustic sensors. A microphone is placed at a central position above each drivetrain component at Axle $A1$ and Axle $A2$. The microphones above the motors are depicted as $\textcolor{blue}{\blacksquare}$ and the microphones above the gearboxes are depicted as $\textcolor{cyan}{\blacksquare}$. The positions of the damaged bearings are indicated by red circles ($\textcolor{red}{\bullet}$). The microphone setup at Car B is identical.} 
 \label{fig:waggon_scheme}
 \end{figure}

 Only the first two drivetrains of each car were equipped with microphones. As shown in Fig.~\ref{fig:mic_placement}, a microphone was placed above every drivetrain component by attaching it to the bottom of the railway car. Thus, the distance between drivetrain component and microphone is $\approx$ $30\, \mathrm{cm}$.  The locations of the microphones (depicted as $\textcolor{blue}{\blacksquare}$ and $\textcolor{cyan}{\blacksquare}$) can also be inferred from Fig.~\ref{fig:waggon_scheme}, which shows a schematic view of Car A. Due to their position, the microphones do not only capture the sounds that are emitted by the bearings, but they additionally capture noise that is caused by the railway tracks and other components in the train bogie, e.g., brakes, dampeners, etc. Please note that in the following  the microphones depicted as $\textcolor{blue}{\blacksquare}$ are considered for classification tasks at the motor and that the microphones depicted as $\textcolor{cyan}{\blacksquare}$ are considered for classification tasks at the gearbox.

\subsection{Bearing Damages}\label{sec:bearing_damages}

The investigated bearing damages   are summarized in Tab.~\ref{tab:damgaged_bearings} and are denoted by $A1\_b1$, $A2\_b2$ and $A2\_b3$. The corresponding healthy bearings at Car B are denoted as $B1\_b1$, $B2\_b2$ and $B2\_b3$, respectively. The first identifier $A1$ in $A1\_b1$ refers to the axle, i.e., Axle 1 at Car A, and the second identifier $b1$ refers to bearing $b1$ (cf. Fig.~\ref{fig:waggon_scheme} and Tab.~\ref{tab:damgaged_bearings}). Bearing $A1\_b1$ represents a bearing fault  in a very early stage at the \ac{IR} that was caused by pitting, whereas $A2\_b2$ represents a bearing with a fault  in a slightly more developed stage at the \ac{OR} which was caused by fatigue. Both bearings are located at the \ac{DE} of the motor (M). 
Bearing $A2\_b3$ is installed at the \ac{NDE} of the gearbox (G) and  represents a fault in a developed stage. This fault is located in the \ac{OR} and was caused due to fatigue. The damaged bearings at the motor are deep groove ball bearings, whereas the bearing at the gearbox is a  cylindrical roller bearing. Note that the considered bearing damages did not develop naturally but were introduced artificially.
\subsection{Data Acquisition}\label{sec:data_acquisition}
 Omni-directional electret microphones of the type  'M 370' were used to capture the signals with a sampling frequency of $25.6 \,\mathrm{kHz}$. The acquired signals were framed into non-overlapping frames with length of 2048 samples. Further, only frames with a mean rotational frequency  $ 42\,\mathrm{Hz} \leq \bar{f}_r \leq 45 \mathrm{Hz}$ of the axle were considered in our evaluatuion since this was the most-used rotational frequency range during  the measurement. No restrictions were made w.r.t. to the applied torque, the mode of the power converter, or the driving direction. Consequently, roughly $28000$ frames were obtained for each microphone. 
 \begin{table}[htb!]
  \caption{Description of the damaged bearings at the first two axles of Car A, i.e., $A1$ and $A2$ in the field measurements. OR: Outer ring fault, IR: Inner ring fault, DE: Drive-end, NDE: Non-drive end, M: Motor, G:Gearbox.}
    \centering
    \resizebox{\columnwidth}{!}{%
    \begin{tabular}{c l l l l l}
        Bearing & Bearing Type & Damage & Location & Axle & Description     \\
        \toprule
         $A1\_b1$ &  deep groove ball bearing  & IR & DE (M) & $A1$ & Pitting damage \\
         $A2\_b2$ &  deep groove
ball bearing  & OR  & DE (M) & $A2$ & Fatigue damage \\
         $A2\_b3$ &   cylindrical roller bearing & OR & NDE (G) & $A2$ & Fatigue damage \\         \bottomrule
    \end{tabular}
    }
   
    \label{tab:damgaged_bearings}
\end{table}
\section{Bearing Fault Detection Approach}\label{sec:approach}
In the following the bearing fault detection approach based on the analysis of airborne sound is presented.
\subsection{Airborne sound for the Detection of Bearing Faults}
\begin{figure}[ht!]
    \centering
     \setlength{\fheight}{14.5cm}
   \setlength{\fwidth}{0.8\columnwidth}
    \input{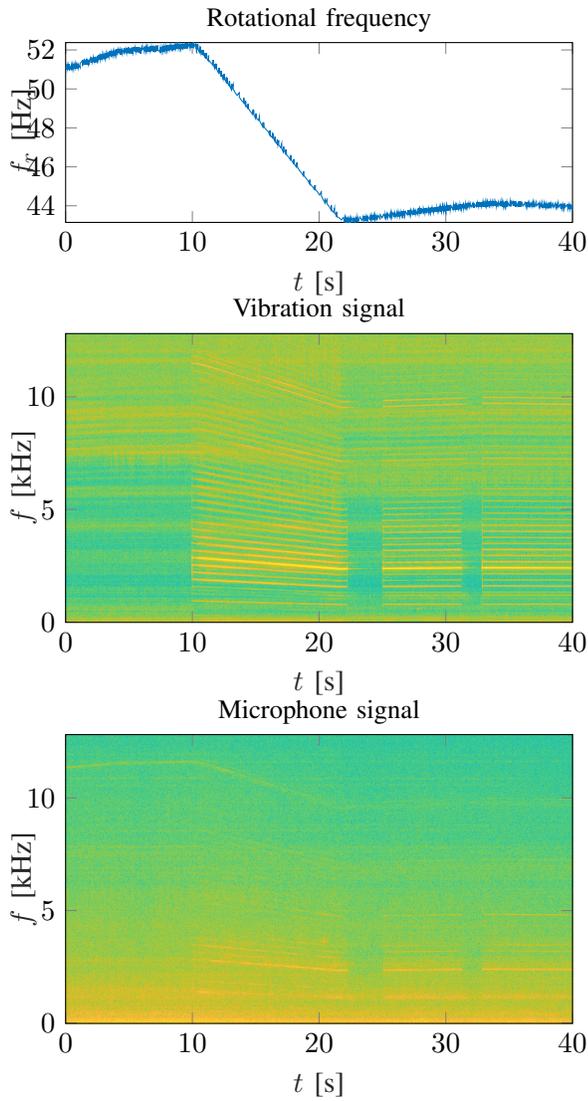}
    \vspace{-0.2cm}
    \caption{Rotational frequency (top), spectrogram of the vibration signal (center) and spectrogram of the microphone signal (bottom) of a healthy bearing.}
    \label{fig:spec_comparison_vib}
\end{figure}
     At first, we want to illustrate the suitability of airborne sound for the task of fault detection by comparing spectrograms obtained from airborne and structure-borne sound signals. Fig.~\ref{fig:spec_comparison_vib} shows a rotational frequency curve (top figure) and the corresponding spectrograms for the structure-borne sound measured directly on the housing of the motor of Axle $B1$ at Car B (center figure) and the spectrogram for the airborne sound (bottom figure) that was captured by the microphone located at the motor at the same axle and at the same car, i.e., $B1\_b1$.  The considered signal shows a period of an almost constant rotational speed which is followed by another period of constant rotational speed after a deceleration period of approximately $10\,\mathrm{s}$. The spectrogram of a vibration signal is dominated by harmonics of the rotational frequency. These harmonics can be clearly observed in the spectrum over the entire frequency range for time periods when torque is applied. If we now compare this to the spectrogram of the microphone signal, a similar structure can be observed, although much less pronounced. While for the vibration signal the harmonics of the rotational frequency could be observed as sharp horizontal lines over the whole frequency range, for the microphone signal only a few harmonics can be faintly recognized  and only in the frequency region below $5\, \mathrm{kHz}$. Thus, it can be stated that spectrograms computed for airborne sound signals exhibit a similar structure as the spectrograms for structure-borne sound signals, although the components that are linked to periodic events in the motor are represented incompletely and less pronounced.
    Consequently, harmonics from bearings are recognizable in the sound signal that is captured in its proximity and thus changes w.r.t the condition of bearing should be detectable using adequate features.
\subsection{Mel-Frequency Cepstral Coefficients}
    In \cite{kreuzer2021} it was shown that bearing faults could be reliably detected with state-of-the-art features for acoustic scene classification tasks that were extracted from structure-borne sound data. More specifically, the first 13 \acp{MFCC} were computed for vibration signals and were used as features to train a One-Class SVM where accuracies above $96\%$ could be obtained for laboratory data. \acp{MFCC} are state-of-the-art features for acoustic scene classification and speaker recognition tasks \cite{schafer2007} as they allow for a compact representation of the spectrum of a signal by combining the cepstrum with a scaling of the frequency on the Mel-Scale. 
    In order to obtain the \acp{MFCC} the following computation steps are required. First, the time domain signal is split into frames by windowing. Then, an $N$-point \ac{DFT} is computed of the windowed input samples. The obtained power magnitude spectrum is then filtered using a Mel-filterbank. Finally, the \acp{MFCC} are obtained by computing the \ac{DCT} of the logarithm of summed filter bank energies. For time-frame $n$, the spectral energies $X_{i}[n]$ of time frame $i$ with $i =0,\ldots, K-1$ can be computed as 
        \begin{align}
        X_i[n] = \sum_{\nu=0}^{N-1} g_{i\nu} \left| \sum_{k=0}^{N-1}  x[n-k] w[k] e^{-\frac{2\pi k\nu}{N}}\right|^2,
    \end{align} 
where $x[k]$ are the time-domain input samples, $w[k]$ is a window function and $g_{i\nu} $ are the samples of a triangular window sequence for weighting the $\nu$-th frequency bin for the $i$-th channel of the  Mel filterbank output, denoted by $X_i[n]$. In logarithmic form, these outputs are then transformed by  a \ac{DCT} to the cepstral domain to yield the \acp{MFCC}s $c_{\mu}$:
\begin{align}
    c_\mu = \sum_{i=1}^{K} \log X_{i}[n] \cos\left( \frac{\pi (2i-1)\mu} {2K} \right), \; \mu = 1, \ldots , K.
\end{align}

\begin{figure}[htb!]
    \centering                 
    \setlength{\fheight}{5cm}
   \setlength{\fwidth}{4.5cm}
    \input{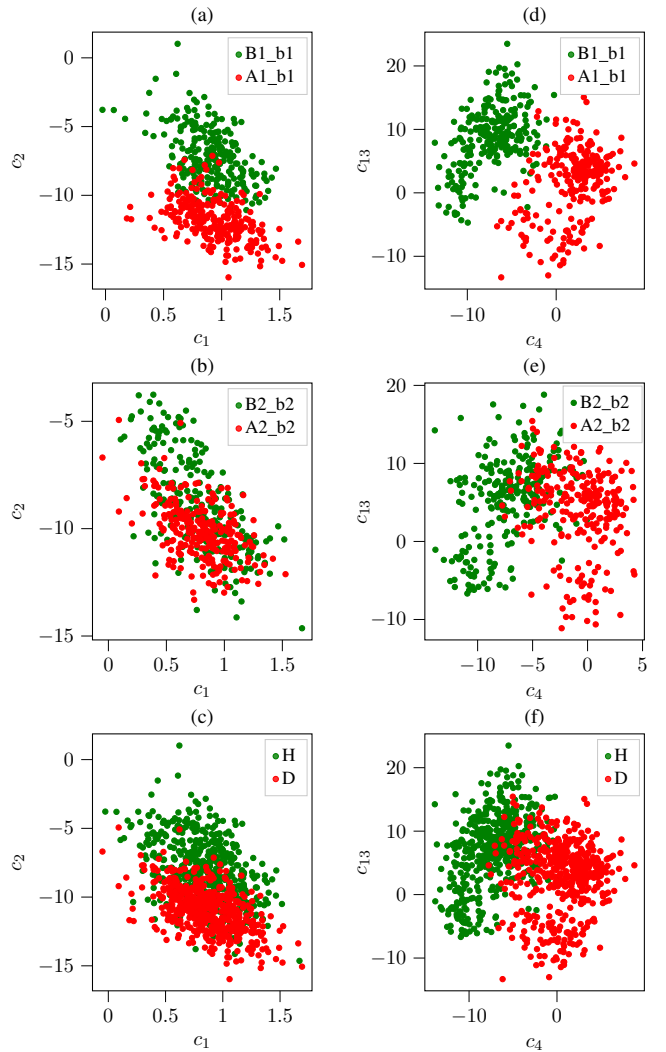}
    \vspace{-0.2cm}
    \caption{\ac{MFCC} scatter plots for the microphone data of the motor. Figures (a)-(c) show scatter plots for the values the first \ac{MFCC}, i.e., $c_{1}$,   on the x-axis and the values for the 2nd  \ac{MFCC}, i.e., $c_{2}$, on y-axis, whereas Figures (d)-(f) show scatter plots for 4th \ac{MFCC}, i.e., $c_{4}$, and the 13th \ac{MFCC}, i.e, $c_{13}$. For figures (c) and (f) the data points for $B1\_b1$ and $B2\_b2$ and $A2\_b2$ and $A1\_b1$ were combined to H and D, respectively.}
    \label{fig:mfcc_scatter}
\end{figure}

    Motivated by the results for structure-borne sound, we investigate in this paper whether \acf{MFCC} are also useful features for the classification bearing faults by analyzing airborne sound. An indication of their appropriateness is given by the following example: Fig.~\ref{fig:mfcc_scatter} shows two-dimensional scatter plots for two exemplary combinations of \acp{MFCC}. Subfigures (a) - (c) in Fig.~\ref{fig:mfcc_scatter} depict scatter plots in which the values of the first \ac{MFCC}, i.e., $c_1$, are shown along the x-axis and the values for the second  \ac{MFCC}, i.e., $c_2$,  are shown on the y-axis. Subfigure (a) shows the data points for the healthy bearing $B1\_b1$ and the damaged bearing $A1\_b1$, respectively, whereas Subfigure (b) shows the data points for $B2\_b2$ and $A2\_b2$. $B1\_b1$ and $B2\_b2$ denote the data that was obtained from the microphones above the motors at Axle $B1$ and Axle $B2$ of Car B, respectively (cf. Fig.~\ref{fig:train_scheme}). For Subfigure (c) the data points for $B1\_b1$ and $B2\_b2$ and $A2\_b2$ and $A1\_b1$ are combined for the labels H (Healthy) and D (Damaged), respectively.  For Subfigures (d) - (f) a different combination of \acp{MFCC} is used: the values for the fourth \ac{MFCC}, i.e, $c_4$, are given along the x-axis and the values of the thirteenth \ac{MFCC}, i.e, $c_{13}$, are plotted on the y-axis. Again,  the data points for  $B1\_b1$ and $B2\_b2$  and $A1\_b1$ and $A2\_b2$ are combined for the labels H and D, respectively,  in Subfigure (f).
    Subfigures (a)-(f) in Fig.~\ref{fig:mfcc_scatter} show that the point clouds for the two classes, Healthy (H) and Damaged (D), do only overlap to a small extent in these two exemplary 2-dimensional subspaces. Since it is already possible to roughly discriminate between healthy and damaged samples in these 2-dimensional subspaces, it is reasonable to assume that the point clouds in the 13-dimensional subspace that is spanned by all 13 \acp{MFCC} are more distant.

\subsection{Classification}\label{sec:classification}
 
     The first 13 \ac{MFCC} are extracted from the raw audio signals and are used as features for training a classifier implemented as a feed forward \ac{ANN} with 2 fully-connected hidden layers of size 1024 and 100, respectively. The number of neurons in the input layer corresponds to the number of features, i.e, 13 in this case, and the number of neurons in the output layer depends on the  number of classes, i.e, two. The output layer is followed by a Softmax activation \cite{Goodfellow2016}. The weights of the network are optimized by applying a cross-entropy loss \cite{Goodfellow2016} and the Adam \cite{Goodfellow2016} optimization algorithm. For a more in-depth treatment of the elements of such an ANN it is referred to \cite{bishop2006,Goodfellow2016}.
   
\section{Results}\label{sec:results}
In our evaluation we distinguish between scenarios with seen bearing damages and unseen bearing damages.  At first, we consider binary classification
tasks with seen bearing damages for both drivetrain components. To this end, the classification of bearing damages at the motor and the gearbox are investigated in Sec.~\ref{sec:damages_engine} and Sec.~\ref{sec:damages_gearbox}, respectively. Then, more challenging and more practical relevant experiments with unseen damages are conducted for the motor in Sec.~\ref{sec:unseen_damages}.
\subsection{Experiments with Seen Damages}\label{sec:seen_damages}
In order to evaluate the performance w.r.t. seen damages we consider binary classifications tasks at the motor in Sec.~\ref{sec:damages_engine} and at the gearbox in Sec.~\ref{sec:damages_gearbox}.
\subsubsection{Binary Classification for Bearing Damages at the Motor}\label{sec:damages_engine}
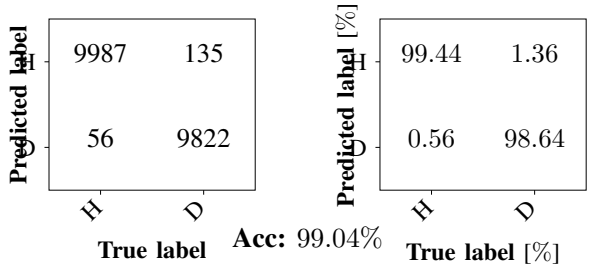
\begin{figure}[ht!]
    \centering
    \begin{tikzpicture}[scale=0.4,every node/.style={scale=2.1}]

\begin{axis}[
tick align=outside,
tick pos=left,
x grid style={white!69.0196078431373!black},
xlabel={\textbf{True label}},
xmin=-0.5, xmax=1.5,
xtick style={color=black},
xtick={0,1},
xticklabel style={rotate=45.0,yshift=0.2cm},
xticklabels={H, D},
y dir=reverse,
y grid style={white!69.0196078431373!black},
ylabel={\textbf{Predicted label} },
ylabel style={yshift=-0.4cm},
ymin=-0.5, ymax=1.5,
ytick style={color=black},
ytick={0,1},
yticklabels={H, D}
]

\draw (axis cs:0,0) node[
  scale=1.1,
  anchor=base,
  text=black,
  rotate=0.0
]{9987};
\draw (axis cs:1,0) node[
  scale=1.1,
  anchor=base,
  text=black,
  rotate=0.0
]{135};

\draw (axis cs:0,1) node[
  scale=1.1,
  anchor=base,
  text=black,
  rotate=0.0
]{56};
\draw (axis cs:1,1) node[
  scale=1.1,
  anchor=base,
  text=black,
  rotate=0.0
]{9822};

\end{axis}
\node[scale=0.5] at (8.6,-1.58) {\small \textbf{Acc: $99.04\%$}};

\begin{scope}[shift={(11,0)}]
\begin{axis}[
tick align=outside,
tick pos=left,
x grid style={white!69.0196078431373!black},
xlabel={\textbf{True label} [$\%$]},
xmin=-0.5, xmax=1.5,
xtick style={color=black},
xtick={0,1},
xticklabel style={rotate=45.0,yshift=0.2cm},
xticklabels={H,D},
y dir=reverse,
y grid style={white!69.0196078431373!black},
ylabel={\textbf{Predicted label} [$\%$]},
ylabel style={yshift=-0.4cm},
ymin=-0.5, ymax=1.5,
ytick style={color=black},
ytick={0,1},
yticklabels={H,D}
]

\draw (axis cs:0,0) node[
  scale=1.1,
  anchor=base,
  text=black,
  rotate=0.0
]{$99.44 $};
\draw (axis cs:0,1) node[
  scale=1.1,
  anchor=base,
  text=black,
  rotate=0.0
]{$0.56 $};

\draw (axis cs:1,0) node[
  scale=1.1,
  anchor=base,
  text=black,
  rotate=0.0
]{$1.36$};
\draw (axis cs:1,1) node[
  scale=1.1,
  anchor=base,
  text=black,
  rotate=0.0
]{$98.64 $};

\end{axis}
\end{scope}
\end{tikzpicture}
    \caption{Confusion matrix for a binary classification problem at the motor with seen
damages.}
\vspace{-0.2cm}
    \label{fig:engine_seen_damages}
\end{figure}

For the first classification task the classifier is trained with data from the first two axles of Car A, i.e., $A1\_b1$ and $A1\_b2$,  and Car B, i.e., $B1\_b1$ and $B1\_b2$. The data from Car A is labelled as Damaged (D) whereas the data recorded at Car B is labelled as Healthy (H). For training and testing the data was split randomly into datasets of sizes 80000 and 20000 samples, respectively. The classifier is trained for 20 epochs with a learning rate of $\lambda=0.001$ and a batch size of 32. The results are summarized in Fig.~\ref{fig:engine_seen_damages}. It can be observed that the bearing faults can be very well detected with a \ac{TPR} of $98.64\%$, while the probability of false alarms is negligibly small with $0.56\%$. Note that for the computation of \acf{TPR} and \ac{TNR} the class label H refers to the positive class and D to the negative class.

\subsubsection{Binary Classification for Bearing Damages at the Gearbox}\label{sec:damages_gearbox}
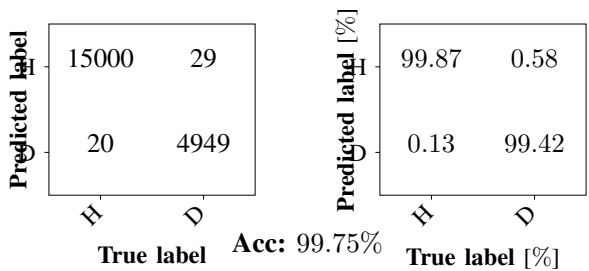
\begin{figure}[ht!]
    \centering
    \begin{tikzpicture}[scale=0.4,every node/.style={scale=2.1}]

\begin{axis}[
tick align=outside,
tick pos=left,
x grid style={white!69.0196078431373!black},
xlabel={\textbf{True label}},
xmin=-0.5, xmax=1.5,
xtick style={color=black},
xtick={0,1},
xticklabel style={rotate=45.0,yshift=0.2cm},
xticklabels={H, D},
y dir=reverse,
y grid style={white!69.0196078431373!black},
ylabel={\textbf{Predicted label} },
ylabel style={yshift=-0.4cm},
ymin=-0.5, ymax=1.5,
ytick style={color=black},
ytick={0,1},
yticklabels={H, D}
]

\draw (axis cs:0,0) node[
  scale=1.1,
  anchor=base,
  text=black,
  rotate=0.0
]{15000};
\draw (axis cs:1,0) node[
  scale=1.1,
  anchor=base,
  text=black,
  rotate=0.0
]{29};

\draw (axis cs:0,1) node[
  scale=1.1,
  anchor=base,
  text=black,
  rotate=0.0
]{20};
\draw (axis cs:1,1) node[
  scale=1.1,
  anchor=base,
  text=black,
  rotate=0.0
]{4949};

\end{axis}
\node[scale=0.5] at (8.6,-1.58) {\small \textbf{Acc: $99.75\%$}};

\begin{scope}[shift={(11,0)}]
\begin{axis}[
tick align=outside,
tick pos=left,
x grid style={white!69.0196078431373!black},
xlabel={\textbf{True label} [$\%$]},
xmin=-0.5, xmax=1.5,
xtick style={color=black},
xtick={0,1},
xticklabel style={rotate=45.0,yshift=0.2cm},
xticklabels={H,D},
y dir=reverse,
y grid style={white!69.0196078431373!black},
ylabel={\textbf{Predicted label} [$\%$]},
ylabel style={yshift=-0.4cm},
ymin=-0.5, ymax=1.5,
ytick style={color=black},
ytick={0,1},
yticklabels={H,D}
]

\draw (axis cs:0,0) node[
  scale=1.1,
  anchor=base,
  text=black,
  rotate=0.0
]{$99.87 $};
\draw (axis cs:0,1) node[
  scale=1.1,
  anchor=base,
  text=black,
  rotate=0.0
]{$0.13 $};

\draw (axis cs:1,0) node[
  scale=1.1,
  anchor=base,
  text=black,
  rotate=0.0
]{$0.58 $};
\draw (axis cs:1,1) node[
  scale=1.1,
  anchor=base,
  text=black,
  rotate=0.0
]{$99.42 $};

\end{axis}
\end{scope}
\end{tikzpicture}
    \vspace{-0.2cm}
    \caption{Confusion matrix for a binary classification problem at the gearbox with seen
damages.}
    \label{fig:gearbox_seen_damages}
\end{figure}

A similar experiment is conducted for the gearbox. Since only a single bearing damage, i.e, $A2\_b3$,  for the gearbox is available, the data which is obtained for the healthy gearboxes at the three remaining axles, i.e, axles $B1$, $B2$ and $A1$ are labelled as Healthy (H). The size of the test set and training set are identical to the previous experiment. The obtained confusion matrices are shown in Fig.~\ref{fig:gearbox_seen_damages}. Again, an almost perfect classification result can be achieved with an accuracy of $99.75\%$. However, it has to be stated at this point that it cannot be ruled out  that the bearing fault at the motor $A2\_b2$ affects the microphone signal that is recorded at the microphone above the gearbox.

\subsection{Experiments with Unseen Damages}\label{sec:unseen_damages}
Since the previous experiments have demonstrated that the detection of bearing faults works
very well for seen data, it is now examined how well this classifier performs for unseen data.
To this end, the classifier is trained with data from one of the two damaged bearings at the
motor, either $A1\_b1$ or $A2\_b2$, and data from the corresponding axle at Car B, i.e., $B1\_b1$ or $B2\_b2$,  and tested with data from the remaining damaged bearing and its corresponding healthy reference. Hence, if the classifier is trained with $A1\_b1$ and $B1\_b1$, it is tested with $A2\_b2$ and $B2\_b2$ and vice-versa.
The results are shown in Figs. \ref{fig:unseen_damage_B2}-\ref{fig:unseen_damage_B1}. Here, the training set and the test set contained
50000 samples each. It can be observed that the overall accuracy drops below $93\%$ in both scenarios. It is notable that an almost inverse behavior of \ac{TPR} and \ac{TNR} can be observed when Fig. \ref{fig:unseen_damage_B2} and Fig.~\ref{fig:unseen_damage_B1} are compared: While a \ac{TPR} of $87.40\%$ and  a \ac{TNR} of $97.46\%$ are obtained when the classifier is tested with data from $A1\_b1$ and its healthy reference, a \ac{TPR} of $98.62\%$ and  a \ac{TNR} of $87.22\%$ is obtained when the classifier is tested with data from $A2\_b2$ and its healthy reference. A possible explanation for this is that while Bearing $A2\_b2$ represents a fault in a developed stage, $A1\_b1$ represents a bearing in a very early stage. Hence, it is somehow plausible that, when the classifier was only trained with a developed fault,  the number of undetected faults increases when it is tested with data from a less severe fault.
The experiments with unseen data support the claim that the features are indeed
fault-related and not axle-related, since the bearing faults at the motor could still be reliably detected  although the faults are positioned at different axles.

In summary, it can be stated that bearing faults can be very well detected with acoustic
data and a feature-based classification approach. However, it should be noted that the
available datasets only contained a small number of bearing faults, which increases the risk of overfitting when training the ANN.
\begin{figure}[hbt!]
    \centering
    \begin{tikzpicture}[scale=0.4,every node/.style={scale=2.1}]

\begin{axis}[
tick align=outside,
tick pos=left,
x grid style={white!69.0196078431373!black},
xlabel={\textbf{True label}},
xmin=-0.5, xmax=1.5,
xtick style={color=black},
xtick={0,1},
xticklabel style={rotate=45.0,yshift=0.2cm},
xticklabels={H, D},
y dir=reverse,
y grid style={white!69.0196078431373!black},
ylabel={\textbf{Predicted label} },
ylabel style={yshift=-0.4cm},
ymin=-0.5, ymax=1.5,
ytick style={color=black},
ytick={0,1},
yticklabels={H, D}
]

\draw (axis cs:0,0) node[
  scale=1.1,
  anchor=base,
  text=black,
  rotate=0.0
]{21850};
\draw (axis cs:0,1) node[
  scale=1.1,
  anchor=base,
  text=black,
  rotate=0.0
]{3150};

\draw (axis cs:1,0) node[
  scale=1.1,
  anchor=base,
  text=black,
  rotate=0.0
]{636};
\draw (axis cs:1,1) node[
  scale=1.1,
  anchor=base,
  text=black,
  rotate=0.0
]{24363};

\end{axis}
\node[scale=0.5] at (8.6,-1.58) {\small \textbf{Acc: $92.43\%$}};

\begin{scope}[shift={(11,0)}]
\begin{axis}[
tick align=outside,
tick pos=left,
x grid style={white!69.0196078431373!black},
xlabel={\textbf{True label} [$\%$]},
xmin=-0.5, xmax=1.5,
xtick style={color=black},
xtick={0,1},
xticklabel style={rotate=45.0,yshift=0.2cm},
xticklabels={H,D},
y dir=reverse,
y grid style={white!69.0196078431373!black},
ylabel={\textbf{Predicted label} [$\%$]},
ylabel style={yshift=-0.4cm},
ymin=-0.5, ymax=1.5,
ytick style={color=black},
ytick={0,1},
yticklabels={H,D}
]

\draw (axis cs:0,0) node[
  scale=1.1,
  anchor=base,
  text=black,
  rotate=0.0
]{$87.40 $};
\draw (axis cs:0,1) node[
  scale=1.1,
  anchor=base,
  text=black,
  rotate=0.0
]{$12.60 $};

\draw (axis cs:1,0) node[
  scale=1.1,
  anchor=base,
  text=black,
  rotate=0.0
]{$2.54 $};
\draw (axis cs:1,1) node[
  scale=1.1,
  anchor=base,
  text=black,
  rotate=0.0
]{$97.46 $};

\end{axis}
\end{scope}
\end{tikzpicture}
    \vspace{-0.2cm}
    \caption{Confusion matrix for a binary classification problem with the unseen Bearing Damage $A2\_b2$ in the test set.}
    \label{fig:unseen_damage_B2}
\end{figure}
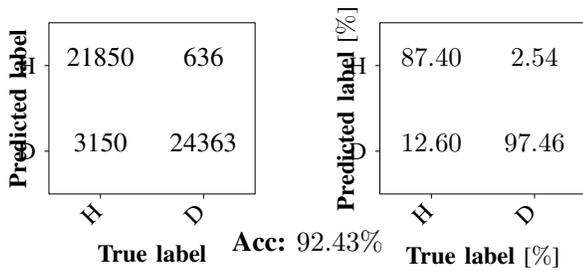
\begin{figure}[hbt!]
    \centering
    \begin{tikzpicture}[scale=0.4,every node/.style={scale=2.1}]

\begin{axis}[
tick align=outside,
tick pos=left,
x grid style={white!69.0196078431373!black},
xlabel={\textbf{True label}},
xmin=-0.5, xmax=1.5,
xtick style={color=black},
xtick={0,1},
xticklabel style={rotate=45.0,yshift=0.2cm},
xticklabels={H, D},
y dir=reverse,
y grid style={white!69.0196078431373!black},
ylabel={\textbf{Predicted label} },
ylabel style={yshift=-0.4cm},
ymin=-0.5, ymax=1.5,
ytick style={color=black},
ytick={0,1},
yticklabels={H, D}
]

\draw (axis cs:0,0) node[
  scale=1.1,
  anchor=base,
  text=black,
  rotate=0.0
]{24656};
\draw (axis cs:0,1) node[
  scale=1.1,
  anchor=base,
  text=black,
  rotate=0.0
]{344};

\draw (axis cs:1,0) node[
  scale=1.1,
  anchor=base,
  text=black,
  rotate=0.0
]{3194};
\draw (axis cs:1,1) node[
  scale=1.1,
  anchor=base,
  text=black,
  rotate=0.0
]{21805};

\end{axis}
\node[scale=0.5] at (8.6,-1.58) {\small \textbf{Acc: $92.92\%$}};

\begin{scope}[shift={(11,0)}]
\begin{axis}[
tick align=outside,
tick pos=left,
x grid style={white!69.0196078431373!black},
xlabel={\textbf{True label} [$\%$]},
xmin=-0.5, xmax=1.5,
xtick style={color=black},
xtick={0,1},
xticklabel style={rotate=45.0,yshift=0.2cm},
xticklabels={H,D},
y dir=reverse,
y grid style={white!69.0196078431373!black},
ylabel={\textbf{Predicted label} [$\%$]},
ylabel style={yshift=-0.4cm},
ymin=-0.5, ymax=1.5,
ytick style={color=black},
ytick={0,1},
yticklabels={H,D}
]

\draw (axis cs:0,0) node[
  scale=1.1,
  anchor=base,
  text=black,
  rotate=0.0
]{$98.62 $};
\draw (axis cs:0,1) node[
  scale=1.1,
  anchor=base,
  text=black,
  rotate=0.0
]{$1.38 $};

\draw (axis cs:1,0) node[
  scale=1.1,
  anchor=base,
  text=black,
  rotate=0.0
]{$12.78 $};
\draw (axis cs:1,1) node[
  scale=1.1,
  anchor=base,
  text=black,
  rotate=0.0
]{$87.22$};

\end{axis}
\end{scope}
\end{tikzpicture}
    \vspace{-0.2cm}
    \caption{Confusion matrix for a binary classification problem with the unseen Bearing Damage $A1\_b1$ in the test set.}
    \label{fig:unseen_damage_B1}
\end{figure}
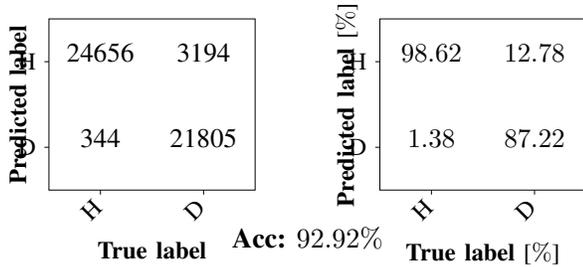

\section{Conclusion}\label{sec:conclusion}
In this paper it was demonstrated that the classification of bearing faults in railway vehicles using airborne sound data in the field is possible even in a realistic and thus highly challenging scenario. For this, acoustic features, i.e., \acp{MFCC}, were extracted from sound signals that were not recorded on a test bench but on a rail vehicle during regular operation on the railway network. Almost perfect classification results can be obtained for scenarios with seen damages, and for scenarios with unseen data accuracies above $92\%$ could still be obtained. In summary, it can be stated that airborne sound data is well-suited for the detection and classification of bearing faults. Consequently, microphones can be a valuable addition to acceleration sensors and should hence be investigated further. The \acp{MFCC} are an empirical choice and its parameters were optimized for a small number of fault types and a single scenario. To this end, the promising results should be confirmed in other scenarios, e.g., the detection of rotor imbalance, and for more complex classification tasks, e.g., multi-class classification with a variety of bearing damages, which would however require more than the hitherto available data. Further, it is worth investigating whether a bi-modal approach, i.e., the combination of structure-borne and airborne sound for classification, can lead to even better results.


\bibliographystyle{IEEEtran}
\bibliography{bib}
\end{document}